# Mechanics of Adhered, Pressurized Graphene Blisters

Narasimha G. Boddeti[1], Steven P. Koenig[1], Rong Long[1,2], Jianliang Xiao[1], J. Scott Bunch[1], and Martin L. Dunn[3]

[1]*Department of Mechanical Engineering*
*University of Colorado at Boulder*
*Boulder, Colorado 80309*

[2]*Current address:*
*Department of Mechanical Engineering*
*University of Alberta*
*Edmonton, Alberta T6G 2G8, Canada*

[3]*Singapore University of Technology and Design*
*Singapore, 138682*

**Abstract**
We study the mechanics of pressurized graphene membranes using an experimental configuration that allows the determination of the elasticity of graphene and the adhesion energy between a substrate and a graphene (or other two-dimensional solid) membrane. The test consists of a monolayer graphene membrane adhered to a substrate by surface forces. The substrate is patterned with etched microcavities of a prescribed volume and when they are covered with the graphene monolayer it traps a fixed number (N) of gas molecules in the microchamber. By lowering the ambient pressure, and thus changing the pressure difference across the graphene membrane, the membrane can be made to bulge and delaminate in a stable manner from the substrate. This is in contrast to the more common scenario of a constant pressure membrane blister test where membrane delamination is unstable and so this is not an appealing test to determine adhesion energy. Here we describe the analysis of the membrane/substrate as a thermodynamic system and explore the behavior of the system over representative experimentally-accessible geometry and loading parameters. We carry out companion experiments and compare them to the theoretical predictions and then use the theory and experiments together to determine the adhesion energy of graphene/$SiO_2$ interfaces. We find an average adhesion energy of 0.24 $J/m^2$ which is lower, but in line with our previously reported values. We assert that this test – which we call the constant N blister test – is a valuable approach to determine the adhesion energy between two-dimensional solid membranes and a substrate, which is an important, but not well-understood aspect of behavior. The test also provides valuable information that can serve as the basis for subsequent research to understand the mechanisms contributing to the observed adhesion energy. Finally, we show how in the limit of a large microcavity, the constant N test approaches the behavior observed in a constant pressure blister test and we provide an experimental observation that suggests this behavior.

## 1. Introduction

Graphene consists of a single, or a few, layers of carbon atoms bonded by strong covalent bonds within a layer, but weaker van der Waals bonds between layers. A monolayer of graphene represents the ultimate limit in thickness for two-dimensional solids. Graphene has impressive electrical, physical, and mechanical properties (Geim, 2009) and as a result has been pursued for many technological applications including electronics, barriers, and energy storage (Lin et al., 2010; Chen et al., 2011; El-kady et al., 2012; Zhu et al., 2012). Because graphene is so thin, it can also be extremely compliant when it has in-plane dimensions on the order of only a few microns and this makes structures fabricated from graphene susceptible to adhesion to a substrate or neighboring structures.

Myriad structures have been created from graphene sheets; some in reality (Low et al., 2012; Scharfenberg et al., 2011; Levy et al., 2010; Pan et al., 2012; Meyer et al., 2007; Kim et al., 2011) and many more in computational simulations (Li and Zhang, 2010; Aitken and Huang, 2010; Lu and Dunn, 2010) that provide important future directions. Blisters are a seemingly simple class of structures that have been observed in various shapes and sizes as a result of graphene fabrication processes and intentionally fabricated to yield attractive technological characteristics, such as strain-engineered electronic properties (Georgiou et al., 2011). Graphene membranes deformed by indentation with an atomic force microscope (AFM) (Lee et al., 2008), intercalation of nanoparticles (Zong et al., 2010), and controlled pressurization by a gas (Bunch et al., 2008; Koenig et al., 2011) have been used to determine various mechanical, and more recently adhesive, properties of graphene. Specifically, in a previous rapid communication (Koenig et al., 2011), we developed a particularly attractive graphene blister test where we mechanically exfoliated graphene membranes (from 1 to 5 layers) on top of a circular cavity (~5 μm diameter and ~300 nm depth) that was microfabricated on a silicon substrate with a thick layer of $SiO_2$ on its surface. This resulted in a graphene membrane adhered to $SiO_2$, presumably by van der Waals forces, and suspended over the cavity with gas trapped inside of it because graphene is impermeable to gas molecules (Bunch et al., 2008; Koenig et al., 2011). We charged this cavity/membrane device in a high-pressure chamber so that the pressure inside and outside the cavity equilibrated at a prescribed value, and then we removed it to ambient at which point the pressure outside the cavity was less than that inside of it and this caused the membrane to bulge. The membrane, or blister, bulged under the condition that a fixed number of molecules of gas was trapped in the chamber. If the charging pressure exceeded a critical value, the blister not only bulged, but it also delaminated from the $SiO_2$ substrate in a stable manner. After delamination the graphene retained the form of a circular blister, but with an increased radius and height and a decreased pressure in the cavity due to the increased volume under the bulge. In our previous communication we used this cavity/blister system, which we termed a *constant N blister test* (N denotes number of molecules) to determine elastic properties of monolayer and multilayer graphene as well as the adhesion energy between graphene and $SiO_2$. Here we describe the mechanics of this test in detail; although it seems fairly straightforward, it admits rich and interesting phenomena across experimentally-accessible system parameters. We demonstrate, through a series of examples, the phenomena of deformation, stability, and interfacial delamination and show how the analysis can be combined with measurements of blister shapes with an AFM to determine elastic and adhesive properties. We use a combination of our previously-reported data and new measurements to demonstrate the utility of these blister tests. In our study of graphene blisters adhered to a substrate we adopt a continuum viewpoint and describe the interaction in terms of an effective adhesion energy that results from the surface forces between the graphene membrane and substrate. We do not consider the origin of these surface forces, e.g., van der Waals, capillary, etc. (DelRio et al., 2007, 2008), which is itself not well understood, and remains a fruitful area for future research.

## 2. Graphene Blisters and the Constant N Pressurized Membrane Test

We consider a blister test structure (Fig. 1) that consists of a circular cylindrical cavity of volume $V_0$ (radius $a_0$) containing $N$ molecules of a gas, an isotropic elastic membrane (Poisson's ratio, $\nu$, Young's modulus, $E$, and thickness, $t$) adhered to the surface of the substrate (adhesion energy, $\Gamma$), and an external environment at a prescribed pressure $p_e$. We realized structures consisting of monolayer graphene membranes adhered to a $SiO_2$ surface through a combination of microfabrication and mechanical exfoliation of graphene. Specifically, we prepared the graphene blisters on two Si wafers, referred to as Chip 1 and Chip 2 hereafter. We photolithographically-defined cylindrical cavities of radii $a_0$ = 2.32 µm and 2.55 µm on Chip 1 and Chip 2, respectively, and the Si surface was thermally oxidized to realize a 285 nm thick layer of $SiO_2$. We etched multiple cylindrical cavities to nominal depths of 293nm and 290nm with reactive ion etching for Chip 1 and Chip 2, respectively. We then deposited suspended graphene sheets over the microcavities via mechanical exfoliation with natural graphite. Our samples consisted of 5 monolayer membranes on Chip 1 and 4 monolayer membranes on Chip 2. We verified that the graphene was a monolayer using a combination of measurement techniques including Raman spectroscopy, optical contrast, AFM, and elastic constants; the procedures are similar to that used in our previous studies (Koenig et al., 2011). The monolayers appear to be quite flat on the substrate, with insignificant pull-in into the cavities.

After exfoliation we *charge* the system in a chamber so that the internal pressure $p_i$ and external pressure are equal at a prescribed value $p_0$. Practically the charging occurs over a period of about seven days as gas molecules ($N_2$ in our study) diffuse through the $SiO_2$ layer and become trapped within the microchamber over the time scale of the remainder of the test. Further details regarding the gas diffusion are given in Bunch et al. (2008) and Koenig et al. (2011).

At this state the membrane is flat, adhered to the substrate at outer perimeter, and spans the cylindrical cavity which holds $N$ gas molecules (Fig. 1a). Removing the device from the chamber has the effect of fixing $p_e$ at a new value $p_e < p_0$, which results in a pressure difference across the membrane that causes it to bulge and increases the volume by $V_b$. Over the time scale of the subsequent measurements diffusion of the gas through the $SiO_2$ is insignificant and so $N$ can be considered fixed; we refer to this as a *Constant N* test as opposed to more common constant pressure membrane inflation tests. As a result, the internal cavity pressure $p_i$ drops to a value $p_i < p_0$. If the charging pressure is below a critical value, $p_{cr}$, the pressure difference $p = p_i - p_e$ across the membrane causes it to bulge into a nearly-spherical cap, while maintaining its adherence to the surrounding substrate. If the charging pressure is greater than $p_{cr}$, the membrane will delaminate from the outer perimeter of the cavity. In the final equilibrium configuration the cavity volume is $V_0 + V_b$ where $V_b$ is the volume of the blister and depends on whether the membrane has delaminated or not.

During our experiments, we use an AFM to measure the shape of the graphene membrane during each stage of the deformation described above. From full-field measurements of the membrane, we extract the maximum deflection, $\delta$, and the blister radius $a$. Initially the radius $a$ is equal to the cavity radius $a_0$, and then becomes larger than $a_0$ due to membrane delamination.

## 3. Analysis of the Blister Test

We model the blister/cavity/substrate configuration as a thermodynamic system with the goal of developing relations among the system parameters (geometry, loading, elastic properties, and the membrane-substrate interface adhesion energy). Our approach is to determine free energy of the thermodynamic system by modelling the gas as ideal and adopting a nonlinear membrane model to describe the deformation of the membrane. We then calculate minimum energy configurations as a function of system parameters and study their stability. In the following we describe the details of this process. Of course our work is related to many other studies of graphene membranes specifically, and membranes more generally and we note specifically that of Yue et al. (2012) that analyzes similar blister configurations and studies the effect of the approximations made in the membrane mechanics, Wan and Mai (1995) who to the best of our knowledge first proposed the blister test with a trapped mass of gas, and Gent and Lewandowski (1987) who analyzed delamination in the constant pressure loading case.

*Mechanics of Pressurized Blisters*
We model the bulged graphene blister as an axisymmetric thin structure clamped at a radial position; before delamination the radial boundary is located at $r = a_o$ and afterwards it is at $r = a$ with $a > a_o$. The mechanical behavior of thin structures can be described by Foppl-von Karman (FvK) plate equations which include contributions from both bending and stretching. For the graphene blisters considered here, we assume that the bending rigidity is negligible and adopt the series solution of the simplified FvK equations obtained by Hencky (1915) that culminates in a relation between the maximum deflection $\delta$, pressure difference across the membrane $p = p_i - p_e$ and the radius of the pressurized circular region $a$:

$$\delta = C_2 \left(\frac{pa^4}{Et}\right)^{\frac{1}{3}} \tag{1}$$

The volume $V_b$ under the bulge is given by:

$$V_b = C_1 \pi a^2 \delta \tag{2}$$

Here $C_1$ and $C_2$ are constants dependent on the Poisson's ratio (it is well-known that $C_1$ and $C_2$ have errors in Hencky's paper, see, e.g., Williams (1987) and Wan and Mai (1995) for corrected versions); we use $C_1 = 0.524$ and $C_2 = 0.687$, consistent with $\nu = 0.16$. Hencky's solution is formally for the case of a uniformly-distributed load on the membrane which simplifies the analysis. Fichter (1997) treated the case of a uniform pressure load on the membrane which is more complicated, but still analytically tractable. For the scenarios considered in this paper, the difference between the uniform load and uniform pressure are small and we neglect them. Furthermore, Hencky's solution does not consider the effects of initial stress in the bulged membrane. Campbell (1956) extended Hencky's solution to cases with an initial tension $N_0 \neq 0$ and showed that when the non-dimensional parameter $P = \frac{pa}{Et}\left(\frac{Et}{N_0}\right)^{\frac{3}{2}} > 100$, the deflection given by eq. (1) is within 5% of the solution with $N_0$ taken into account. Mechanically exfoliated graphene blisters like the ones of our study often have an initial tension, $N_0$ between 0.03-0.15 N/m (e.g., Wang et al., 2012; Bunch et al., 2008; Barton et al., 2011). With typical values of $a = 2$ μm, $Et = 340$ N/m (Lee et al., 2008) and $N_0 = 0.07$ N/m, the non-dimensional parameter $P$ is about 100 when the pressure load is about 500 kPa. The majority of measurements in our experiment are done well above 500 kPa, hence we neglect the effect of $N_0$ and use Hencky's solution to completely describe the mechanics of the pressurized blisters. Nevertheless, the incorporation of $N_0$ is straightforward in practice.

*Thermodynamic Model of the Blister Configuration*
We model the behavior of the blister considering the three stages identified in Fig. 1. Initially the system is at equilibrium with the graphene membrane flat and stress free and the pressure inside and outside the cavity equal to $p_0$ (Fig. 1a). The pressure outside the cavity is then reduced to $p_e$ which causes the membrane to deform due to the pressure difference across it $p = p_i - p_e$. The gas inside the cavity is assumed to isothermally expand to its final equilibrium pressure $p_i$. Depending on the magnitude of $p_e$, one of two configurations will arise: i) the membrane will bulge, but not delaminate (Fig. 1b), or ii) the membrane will both bulge and delaminate (Fig. 1c). In both cases we describe the membrane mechanics using the Hencky solution and parameterize the deformed shape by the radius $a$ and maximum deflection $\delta$; in the former $a = a_0$ and in the latter $a > a_0$.

Our strategy is to determine equilibrium configurations of the deformed membrane by seeking minima in the system free energy $F$. To this end we recognize that the change in free energy of the system can be expressed as:

$$F = F_{mem} + F_{gas} + F_{ext} + F_{adh} \tag{3}$$

In eq. (3), $F_{mem}$ is the strain energy stored in the membrane as it deforms when subjected to a pressure difference across it $p$, $F_{gas}$ is the free energy change associated with expansion of the N gas molecules in the microchamber, $F_{ext}$ is the free energy change of the external environment that is held at a constant pressure $p_e$, and $F_{adh}$ is the adhesion energy of the membrane-substrate interface.

For a fixed $a$, we can compute $F_{mem}$ assuming quasistatic expansion of the gas and using the relations from eqs. (1) and (2):

$$F_{mem} = \iint N_i d\epsilon_i dA_{mem} = \frac{pV_b}{4} \tag{4}$$

where $N_i$ is the membrane force resultant, $\epsilon_i$ is the associated strain, and $dA_{mem}$ is an infinitesimal element of membrane cross sectional area.

The free energy change due to isothermal expansion of the fixed number of gas molecules N in the microchamber, from an initial pressure and volume $(p_0, V_0)$ to final pressure and volume $(p_i, V_0 + V_b)$ is:

$$F_{gas} = -\int P dV = -p_0 V_0 \ln\left[\frac{V_0 + V_b}{V_0}\right] \tag{5}$$

As the blister expands by $V_b$, the volume of the surroundings decreases by an equal amount (assuming no volume change of the membrane). Assuming the surroundings are maintained at a constant pressure $p_e$, the free energy then changes by:

$$F_{ext} = \int p_e dV = p_e V_b \tag{6}$$

For a constant value of adhesion energy per unit area $\Gamma$, $F_{adh}$ is then:

$$F_{adh} = \int \Gamma dA = \Gamma \pi (a^2 - a_0^2) \tag{7}$$

Equations (4) – (7) show that the system energetics are described by three unknowns: $p_i$, $\delta$ and $a$. The constitutive eq. (1) along with the ideal gas equation $p_0 V_0 = p_i(V_0 + V_b)$ provides two relations between these three unknowns; we use these to express the free energy in terms of the single unknown $a$:

$$F(a) = \frac{pV_b}{4} - p_0 V_0 \ln\left[\frac{V_0 + V_b}{V_0}\right] + p_e V_b + \Gamma \pi (a^2 - a_0^2) \tag{8}$$

Recall that $V_b$ is a function of $a$ as given by eqs. (1) and (2). We determine equilibrium configurations by computing extrema of $F(a)$:

$$\frac{dF(a)}{da} = 0 \tag{9}$$

When there is no delamination ($a = a_0$), the equilibrium solution is obtained simply from eqs. (1) and (2) along with the ideal gas equation. When there is delamination, the equilibrium configuration obtained by solving eq. (9) can be expressed as:

$$\frac{dF(a)}{da} = -\frac{3p}{4}\frac{dV_b}{da} + \frac{V_b}{4}\frac{dp}{da} + 2\pi \Gamma a = 0 \tag{10}$$

Here $p$ depends on $a$ through the relation obtained from eq. (1) and ideal gas equation:

$$a = \left(\frac{p_0}{p_i} - 1\right)^{\frac{3}{10}} \left(\frac{V_0}{\pi C_1 C_2}\right)^{\frac{3}{10}} \left(\frac{Et}{p}\right)^{\frac{1}{10}} \tag{11}$$

Using eqs. (1) and (2), we can write:

$$\tag{12}$$

$$\frac{dV_b}{da} = \frac{\partial V_b}{\partial p}\bigg|_a \frac{\partial p}{\partial a} + \frac{\partial V_b}{\partial a}\bigg|_p = \frac{1}{3}\frac{V_b}{p}\frac{\partial p}{\partial a}\bigg|_a + \frac{\partial V_b}{\partial a}\bigg|_p$$

Substituting eq. (12) into eq. (10) results in the relation:

$$\frac{d\mathcal{F}(a)}{da} = -\frac{3p}{4}\frac{\partial V_b}{\partial a}\bigg|_p + 2\pi \Gamma a = 0 \tag{13}$$

Rearranging and using ideal gas equation, we finally obtain:

$$\Gamma = \frac{5C_1}{4}\left(\frac{p_0 V_0}{V_0 + V_b(a)} - p_e\right)\delta(a) \tag{14}$$

Equation (14) describes equilibrium configurations in terms of system parameters $(p_0, p_e, h, a, \delta, \Gamma)$. We use eq. (14) with typical experiments, to determine $\Gamma$ with prescribed values of $p_0$ and $p_e$, $(a, \delta)$ pairs measured with an atomic force microscope, $V_0 = \pi a_0^2 h$ determined by the device geometry and $V_b(a)$ given by eq. (2).

In an experiment if we systematically increase $p_0$, we find that at a critical value, the membrane will begin to delaminate. We determine $p_{cr}$ by substituting $a = a_0$ in eq. (14) and solving for $p_0$:

$$p_{cr} = \left(\left(\frac{4\Gamma}{5C_1 \delta(a_0)}\right) + p_e\right)\frac{V_0 + V_b(a_0)}{V_0} \tag{15}$$

In eq. (15), as $V_0 \to \infty$, $\frac{V_0 + V_b(a_0)}{V_0} \to 1$ and we can express eq. (15) as:

$$\Gamma = \frac{5C_1}{4}(p_{cr} - p_e)\delta(a_0) \tag{16}$$

This agrees with the constant–pressure result obtained by Williams (1997). In essence, as $V_0 \to \infty$, the isothermal expansion approaches a constant pressure process; hence the constant pressure blister configuration results as a limiting case of the constant N blister configuration as the cavity size becomes large.

Finally, we evaluate the stability of the system by computing:

$$\frac{d^2 F}{da^2} = \frac{10 p V_b}{a^2}\left(\frac{2p_0 p_i - 3p_i^2 + p_0 p_e}{3p_0 p + p_i(p_0 - p_i)}\right) \tag{17}$$

If $\frac{d^2F}{da^2} > 0$ the delamination will be stable. Assuming $p_e \ll p_i, p_0$ (which is the case in our experiments) then we require $p_i < 2p_0/3$ for stable delamination. This inequality is equivalent to requiring $V_0 < 2V_b$ which can be satisfied experimentally by tailoring the geometry of the microcavity.

## 4. Results and Discussion

In this section we have three goals: i) to demonstrate the behaviour of the blister system, ii) to use the blister analysis in conjunction with experiments to determine the adhesion energy of graphene-SiO$_2$ interfaces, and iii) to show how the model describes measurements of monolayer graphene blisters in the constant N experimental configurations. Previously (Koenig et al., 2011) we used this blister test to determine the elastic moduli ($Et$) of graphene monolayers and multilayers, but this required more measurements in the elastic regime before delamination than we made here. Since our emphasis here is on the adhesion energy, we did not make as many measurements in the elastic regime and instead used our previous measured modulus results as inputs to our calculations.

*System Behavior: Equilibrium Configurations and Stability*

As mentioned earlier, we obtain equilibrium configurations of the blister system by solving eq. (14) and its stability is described by eq. (17). In general these are implicit equations involving the system parameters, but explicit relations in general are elusive or not particularly revealing so here we describe three specific examples by which we intend to demonstrate the rich behavior of the system for experimentally-accessible system parameters. For each case we prescribe the cavity radius $a_0$ and cavity depth $h$:

Case 1 $a_0 = 2$ μm and $h = 0.25$ μm
Case 2 $a_0 = 3$ μm and $h = 0.25$ μm
Case 3 $a_0 = 2$ μm and $h = 1.25$ μm

For each case we take the membrane to be a graphene monolayer with elastic properties in line with existing measurements and theory $Et = 340$ N/m, $\nu = 0.16$, (Blakslee et al., 1970) and we take $\Gamma = 0.2$ J/m$^2$.

The system in Case 1 has an initial volume $V_0 = \pi a_0^2 h \approx 3.14$ μm$^3$. This geometry is similar to the experimental devices used in our study. From eq. (14) we calculate the critical charging pressure for delamination $p_{cr} = 1.94$ MPa. The free energy of eq. (8) is plotted as a function of the blister radius at three different input/charging pressures as shown in Fig. 2a. The green and magenta colored points on the curves signify the initial configuration of the system and the final equilibrium configuration where $dF/da = 0$ is satisfied respectively. The dashed part of each curve corresponds to $a < a_0$ which is physically not realizable. When $p_0 < p_{cr}$ (blue curve), there is no configuration with free energy less than the initial configuration, implying there will be no delamination and $a$ remains equal to $a_0$. When $p_0 = p_{cr}$ (black curve), the system finds an equilibrium configuration exactly at $a = a_0$, an inflection point. If $p_0$ is increased to a value beyond $p_{cr}$ this unique equilibrium configuration degenerates into two equilibrium configurations –a local maximum with $a < a_0$ (not identified with a symbol and unrealizable) and a local minimum with $a > a_0$ which is evident from the red curve in the Fig. 2a. The presence of this minimum makes the stable delamination possible in the constant N blister test.

From the equilibrium configurations as a function of charging pressure ($p_0$), we obtain various representations of the system behavior. Figure 3a-c shows three quantities as a function of the charging pressure: maximum blister deflection ($\delta$), blister radius ($a$), and cavity pressure ($p_i$). As the charging pressure is increased, the graphene blister deflection increases, and the membrane stiffens resulting in the nonlinear behavior given by eq. (1) and shown by the black curve in Fig. 3a. At $p_0 = p_{cr}$= 1.94 MPa, delamination begins and as $p_0$ continues to increase the blister continues to delaminate and the deflection increases as given by eq. (14). This is shown by the red line in Figs. 3a and 3b, the latter showing the blister radius after the onset of delamination. Figure 3c shows the evolution of the cavity pressure $p_i$ with increasing $p_0$. Before delamination, $p_i$ increases nearly linearly with $p_0$; the gentle softening of the curve results because as the blister volume increases with a constant number of gas molecules trapped in the cavity, the pressure decreases consistent with the ideal gas law. After delamination, $p_i$ decreases rapidly with increasing $p_0$ because the volume increases at higher rate than before delamination, thereby decreasing the equilibrium pressure. Formally, as $p_0 \to \infty$, $p_i \to p_e$.

In the system of Case 2, the radius of the cavity is increased from $a_0 = 2$ μm to 3 μm. In this case, the membrane system is more compliant and, as a result, the critical pressure is lowered from 1.94 MPa to 1.57 MPa. From the $F(a)$ plots in Fig. 2b, at the critical charging pressure the equilibrium now occurs at a minimum rather than at an inflection point. However, this subtle difference from Case 1 does not qualitatively change the system behavior; it behaves similar to that of Case 1 (Fig 3a-c) and so we do not show plots.

Finally in Case 3, we increase the cavity depth $h = 0.25$ to 1.25 μm while keeping the cavity radius at $a_0 = 2$ μm. The critical charging pressure is again decreased from the original 1.94 MPa to 1.39 MPa. The plot of $F(a)$ in Fig. 2c shows that now when $p_0 < p_{cr}$ (blue curve), the curve has two possible extrema instead of none as in the previous two cases.

When the system starts in the prescribed initial configuration, an energy barrier has to be overcome to reach the minimum energy delaminated configuration. When $p_0 = p_{cr}$ (black curve), however, the barrier is removed and the initial configuration coincides with a local maximum. This is an unstable equilibrium and with a small perturbation the system can move to the minimum energy delaminated configuration with $a > a_0$. Therefore, when the charging pressure is increased beyond the critical pressure (1.39 MPa), delamination can occur suddenly with a rapid advance in the membrane radius $a$. This also results in a discontinuity in the equilibrium system parameters as illustrated in Fig. 3d-f. Such a discontinuity is in contrast to the previous two cases where delamination progresses in a stable manner as the charging pressure is increased.

In summary, these case studies show that the equilibrium configuration at the critical charging pressure can be an inflection point, a local minimum, or a local maximum of $F(a)$. What this suggests for experiments is that in the first two cases the blister radius and deflection will evolve as a steady, continuous change from the initial values as the membrane starts delaminating, and similarly the cavity pressure will decrease. In Case 3, however, because the initial condition is an unstable equilibrium there can be a jump in the observable/measured quantities $a$, $\delta$, and $p_i$.

Looking more closely at the behavior we find that as $h$ is increased at a fixed $a_0$, the initial volume $V_0$ can become much larger than the volume of the membrane blister $V_b$. From the ideal gas law for isothermal conditions, $p_i = p_0 V_0 / (V_0 + V_b)$, we see that the pressure $p_i$ approaches the charging pressure $p_0$ when $V_b \ll V_0$. It is well-known that in a constant pressure (P) blister test, delamination is unstable (Gent and Lewandowski, 1987), i.e., once the critical pressure is reached, the entire adhered membrane delaminates. Therefore, for large cavity depths, membrane delamination may initiate in an unstable manner. However, as delamination proceeds, the blister volume $V_b$ increases and eventually becomes comparable to $V_0$. This leads to a significant decrease in the cavity pressure and a stable equilibrium is then approached.

To further illustrate the connection between the constant N and constant P blister tests, we plot the critical pressure versus the cavity depth in Fig. 4a, and see that the constant N blister test curve asymptotically approaches the constant P blister test value which is independent of the cavity depth. Also, the critical pressure as a function of the cavity radius and the adhesion energy is shown in Figs. 4b and 4c, respectively. As the adhesion energy is increased, the critical delamination pressure increases as expected in both constant P and constant N blister tests. While with increasing cavity radius and a fixed cavity depth, the delamination pressure decreases rapidly and continuously in the constant P case whereas in the constant N case it rapidly decreases initially with increasing $a_0$ but reverses this trend after reaching a minimum value.

*Combining the Model and Measurements to Determine Adhesion Energy*
We can determine the adhesion energy $\Gamma$ between the graphene membrane and the substrate (SiO$_2$ in our case) by combining the theory and the experimental measurements. Specifically, using the measured deflection $\delta$ and radius $a$ of the equilibrated blister membrane after delamination (the red symbols in Fig. 7) and the prescribed charging pressure $p_0$, we can calculate the adhesion energy $\Gamma$ from eq. (14). In Figure 5 we plot results obtained in this manner for two different sets of monolayer graphene blisters fabricated on two different chips. The results for Chip 1 are our previously-reported values (Koenig et al., 2011) and show an average adhesion energy of $\Gamma = 0.44$ J/m$^2$. The results shown for Chip 2 are new measurements and show a lower value of $\Gamma = 0.24$ J/m$^2$. The data for both chips are self-consistent, suggesting that the difference is not due to errors in measurements but that it reflects the actual difference in the operant surface forces on the two chips. This in turn could arise from differences in surface properties such as roughness and chemical reactivity, and thus change the apparent adhesion energy. Although the exact cause of the variation in adhesion energies remains to be elucidated with more experimental efforts, these results demonstrate the usefulness of the constant N blister test to determine adhesion energy.

*Blister System Behavior – Measurements and Theory*
Here we compare measurements of monolayer graphene membranes and theory. As mentioned, we used an AFM to measure the deformation of graphene blisters in the constant N configuration. In our measurements we estimate the resolution in blister height to be subnanometerand that in blister radius to be about 90 nm. Figure 6a shows a representative three dimensional profile of a bulged monolayer graphene blister (from Chip 2) and confirms the axisymmetric deformation of the membrane. In Fig. 6b we plot the cross-section of the membrane profile for various values of the prescribed charging pressure $p_0$. When $p_0$ is below 1.32 MPa, the graphene membrane remains attached to the edge of the cavity but as $p_0$ increases the graphene membrane delaminates from the substrate, resulting in a larger radius as shown in Fig. 6b. Also plotted in Fig. 6b are theoretical fits of membrane profiles according to Hencky's solution, with the maximum deflection (eq. (1)) fit to the measurements. The Hencky solution, with the measured maximum deflection as a fitting parameter, is in excellent agreement with the measurements, both in terms of the shape of the profile, but also in terms of the boundary conditions. This reinforces the appropriateness of using Hencky's solution to describe the membrane mechanics in the model of the constant N blister test. In Fig. 7, we plot the measured maximum deflection $\delta$, blister radius $a$ and the calculated equilibrium cavity pressure $p_i$ versus the charging pressure $p_0$ along with theoretical predictions. The behavior is as described in Fig. 3a-c but we also include plots for multiple values of the adhesion energy, centered around the measured value of $\Gamma = 0.24$ J/m$^2$ to illustrate the sensitivity of the measured parameters to the adhesion energy. In Fig. 7 we show the measurements with symbols of two colors, red and black. The black symbols show results before the clear onset of delamination. The red symbols indicate measurements after delamination has occurred and these are used to determine the adhesion energy in Fig. 5. In summary, the theory describes the measurements well.

As we discussed earlier, the theory predicts that when the cavity depth $h$ is large, the blister test system may exhibit an unstable delamination with a jump in the system parameters, including the blister radius. We observed such behavior in tests with microcavities with a cavity radius $a_0 = 2.2$μm and depth $h = 5$μm, a geometry similar to the third example discussed above. We find that with increasing charging pressure $p_0$, graphene membranes bulge as previously described, but that above a critical pressure, the membrane appears to undergo severe delamination with a resulting blister of irregular shape that is very large and covers multiple microcavities; see Fig. 8. In this case, $p_0 = 2.8$ MPa was the pressure at which delamination was observed. We think that this large blister is a consequence of the unstable delamination as predicted by theory and shown in Fig.3c. Conceivably, the membrane delaminated over a large region, neighboring blisters coalesced, and the result is a large irregular shaped blister. Assuming the adhesion energy is between 0.2-0.4 J/m$^2$ and graphene is 8 layered, the predicted critical input pressure for delamination is between 1.90-3.15 MPa. This is in reasonable agreement with the experimental observation where delamination was observed at $p_0 = 2.8$ MPa, but not at a lower pressure of at $p_0 = 2.2$ MPa. We did not do tests at pressures between these two values.

## 5. Conclusions

We studied the mechanics of a graphene membrane adhered to a substrate patterned with etched microcavities of a prescribed volume that trap a fixed number of gas molecules. By lowering the ambient pressure, and thus changing the pressure difference across the graphene membrane, the membrane can be made to bulge and delaminate in a stable manner from the substrate. We analyzed the membrane/substrate as a thermodynamic system and studied the behavior of this the constant N blister test over representative experimentally-accessible geometry and loading parameters. We found that depending on the system parameters, the membrane will deform in a nonlinear elastic manner until a critical charging pressure is reached. At that point, the membrane will delaminate from the substrate in a stable manner. We carried out companion experiments of the membrane deformation as the charging pressure was increased and used them with the theory to determine the adhesion energy of graphene/SiO$_2$ interfaces. We found an average adhesion energy that is lower, but in line with previously reported values by us and others. We also showed that the theoretical predictions described the experiments well, both before and after stable delamination. For deep cavities, the membrane can delaminate in an unstable manner and we demonstrated this experimentally. Although we did not study the nature of the surface forces that influence the adhesion energy, the constant N blister test is an attractive approach to enable the study of important effects on adhesion including substrate topography, membrane stiffness, and the surface force law.

# 6. References

Aitken, Z. H., and Huang, R., 2010, "Effects of mismatch strain and substrate surface corrugation on morphology of supported monolayer graphene," *Journal of Applied Physics*, **107**(12), 123531.

Barton, R. A., Ilic, B., van der Zande, A. M., Whitney, W. S., McEuen, P. L., Parpia, J. M., and Craighead, H. G., 2011, "High, Size-Dependent Quality Factor in an Array of Graphene Mechanical Resonators," *Nano Letters*, **11**(3), pp. 1232-1236.

Blakslee, O. L., Proctor, D. G., Seldin, E. J., Spence, G. B., and Weng, T., 1970, "Elastic constants of compression-annealed pyrolytic graphite," *J. Appl. Phys.*, **41**, pp. 3373–3382.

Bunch, J. S., Verbridge, S. S., Alden, J.S., van der Zande, A. M., Parpia, J. M., Craighead, H. G., and McEuen, P. L., 2008, "Impermeable Atomic Membranes from Graphene Sheets," *Nano Letters*, **8**, pp. 2458-2462.

Bunch, J. S., van der Zande, A. M., Verbridge, S. S., Frank, I. W., Tanenbaum, D. M., Parpia, J. M., Craighead, H. G., and McEuen, P. L., 2007, "Electromechanical Resonators from Graphene Sheets," *Science*, **315**, pp. 490-493.

Campbell, J. D., 1956, "On the theory of initially tensioned circular membranes subjected to uniform pressure," *Q J Mechanics Appl Math*, **9**(1), pp. 84-93.

Chen, S., Brown, L., Levendorf, M., Cai, W., Ju, S-Y., Edgeworth, J., Li, X., Magnuson, C. W., Velamakanni, A., Piner, R. D., Kang, J., Park, J., and Ruoff, R. S., 2011, "Oxidation Resistance of Graphene-Coated Cu and Cu/Ni Alloy," *ACS Nano*, **5**(2), pp. 1321-1327.

DelRio, F., Dunn, M. L., and, de Boer, M. P., 2008, "Capillary Adhesion Model for Contacting Micromachined Surfaces," Scripta Materialia, 59, Viewpoint Set No. 44, pp.916-920.

DelRio, F., Dunn, M. L., Phinney, L. M., Bourdon, C. J., and, de Boer, M. P., 2007, "Rough Surface Adhesion in the Presence of Capillary Condensation," Applied Physics Letters, 90, 163104.

El-Kady, M. F., Strong, V., Dubin, S., and Kaner, R. B., 2012, "Laser Scribing of High-Performance and Flexible Graphene-Based Electrochemical Capacitors," *Science*, **335** (6074), pp. 1326-1330.

Fichter, W. B., 1997, "Some solutions for the large deflections of uniformly loaded circular membranes," *NASA Technical Paper* 3658, Langley Research Center, Hampton, Virginia.

Geim, A. K., 2009, "Graphene: Status and Prospects," *Science,* **324**(5934), pp. 1530-1534.

Gent, A. N., and Lewandowski, L. H., 1987, "Blow-off pressures for adhering layers," *Journal of Applied Polymer Science*, **33**(5), pp. 1567-1577.

Georgiou, T., Britnell, L., Blake, P., Gorbachev, R. V., Gholinia, A., Geim, A. K., Casiraghi, C., and Novoselov, K. S., 2011, "Graphene bubbles with controllable curvature,", *Appl. Phys. Lett.,* **99**, 093103.

Hencky, H., 1915, "Über den spannungszustand in kreisrunden platten mit verschwindender biegungssteiflgkeit," *Z. Fur Mathematik Und Physik,* **63**, pp. 311-317.

Jiang, L. Y., Huang, Y., Jiang, H., Ravichandran, G., Gao, H., Hwang, K. C., and Liu, B., 2006, "A Cohesive Law for Carbon Nanotube/polymer Interfaces Based on the Van Der Waals Force," *J. Mech. Phys. Solids*, **54**, pp. 2436-2452.

Kim, K., Lee, Z., Malone, B. D., Chan, K. T., Alemán, B., Regan, W., Gannett, W., Crommie, M. F., Cohen, M. L., and Zettl, A., 2011, " Multiply folded graphene," *Phys. Rev. B*, **83**, 245433.

Koenig, S. P., Boddeti, N. G., Dunn, M. L., Bunch, J. S., 2011, "Ultrastrong adhesion of graphene membranes," *Nature Nanotechnology,* **6**, pp. 543-546.

Lee, C. G., Wei, X. D., Kysar, J. W., and Hone, J., 2008, "Measurement of the Elastic Properties and Intrinsic Strength of Monolayer Graphene," *Science*, 321, pp. 385-388.

Levy, N., Burke, S. A., Meaker, K. L., Panlasigui, M., Zettl, A., Guinea, F., Castro Neto, A. H., and Crommie, M. F., 2010, "Strain-Induced Pseudo–Magnetic Fields Greater Than 300 Tesla in Graphene Nanobubbles," *Science*, **329**(5991), pp. 544-547.

Li, T., and Zhang, Z., 2010, "Substrate-regulated morphology of graphene," *J. Phys. D: Appl. Phys.*, **43**, 075303.

Lin, Y.-M., Dimitrakopoulos, C., Jenkins, K. A., Farmer, D. B., Chiu, H.-Y., Grill, A., and Avouris, Ph., 2010, "100-GHz Transistors from Wafer-Scale Epitaxial Graphene," *Science*, **327**(5966), pp. 662.

Low, T., Perebeinos, V., Tersoff, J., and Avouris, P., 2012, "Deformation and Scattering in Graphene over Substrate Steps," *Phys. Rev. Lett.*, **108**(9), 096601.

Lu, Z. and Dunn, M. L., 2010, "van der Waals Adhesion of Graphene Membranes," *J. Appl. Phys.*, **96**, 111902.

Meyer, J. C., Geim, A. K., Katsnelson, M. I., Novoselov, K. S., Booth, T. J., and Roth, S., 2007, "The structure of suspended graphene sheets," *Nature*, **446**, pp. 60-63.

Meyer, J. C., Kisielowski, C., Erni, R., Rossell, M. D., Crommie, M. F., and Zettl, A., 2008, "Direct Imaging of Lattice Atoms and Topological Defects in Graphene Membranes," *Nano Lett.*, **8**(11), pp. 3582-3586.

Pan, W., Xiao, J., Zhu, J., Yu, C., Zhang, G., Ni, Z., Watanabe, K., Taniguchi, T., Shi, Y. and Wang, X., 2012, ".Biaxial Compressive Strain Engineering in Graphene/Boron Nitride Heterostructures," *Scientific Reports*, **2**, 893.

Scharfenberg, S., Rocklin, D. Z., Chialvo, C., Weaver, R. L., Goldbart, P. M., and Mason, N., 2011, "Probing the mechanical properties of graphene using a corrugated elastic substrate," *Appl. Phys. Lett.*, **98**(9), 091908.

Springman, R. M., and Bassani, J. L., 2008, "Snap Transitions in Adhesion," *J. Mech. Phys. Solids.*, **56**, pp. 2358-2380.

Tang, T., Jagota, A., and Hui, C. Y., 2005, "Adhesion between Single-walled Carbon Nanotubes," *J. Appl. Phys.*, **97**, 074304.

Wan, K.-T., and Mai, Y.-W., 1995, "Fracture mechanics of a new blister test with stable crack growth," *Acta Metallurgica et Materialia*, **43**(11), pp. 4109-4115.

Wang, L., Travis, J. J., Cavanagh, A. S., Liu, X., Koenig, S. P., Huang, P. Y., George, S. M., and Bunch, J. S., 2012, "Ultrathin Oxide Films by Atomic Layer Deposition on Graphene," *Nano Letters*, **12**(7), pp. 3706-3710.

Williams, J. G., 1997, "Energy Release Rates for the Peeling of Flexible Membranes and the Analysis of Blister Tests," *Int. J. Fracture*, **87**, pp. 265-288.

Yue, K., Gao, W., Huang, R., and Liechti, K. M., 2012, "Analytical methods for mechanics of graphene bubbles,", *J. Appl. Phys.*, **112**(8), 083512.

Zhu, Y., Li, L., Zhang, C., Casillas, G., Sun, Z., Yan, Z., Ruan, G., Peng, Z., Raji, A. O., Kittrell, C., Hauge, R. H., and Tour, J. M., 2012, "A seamless three-dimensional carbon nanotube graphene hybrid material," *Nature Communications*, **3**, 1225.

Zong, Z., Chen, C.-L., Dokmeci, M. R., and Wan, K.-T., 2010, "Direct measurement of graphene adhesion on silicon surface by intercalation of nanoparticles." *J. Appl. Phys.*, **107**(2), 026104.

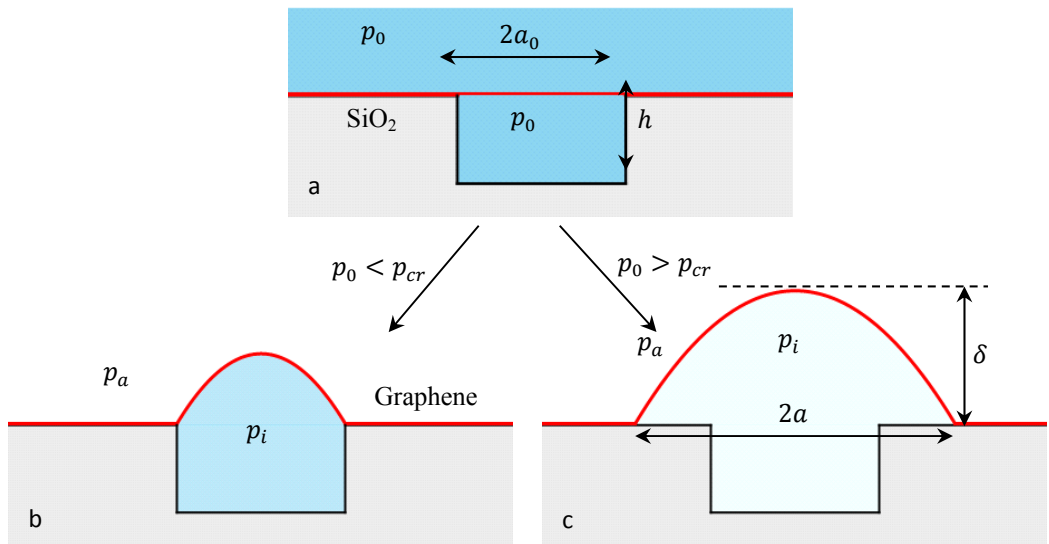

Figure 1  Schematic cross sections of test structures illustrating: (a) the initial configuration of the system, charged to a pressure $p_0$ in a pressure chamber – the blue color indicates gas and the red curve is the graphene membrane; possible final configurations when the external pressure is reduced with graphene membranes deformed due to the expanding gas molecules (b) with; and (c) without delamination from the substrate. The change of the blue color from a darker to a lighter shade indicates decreasing pressure.

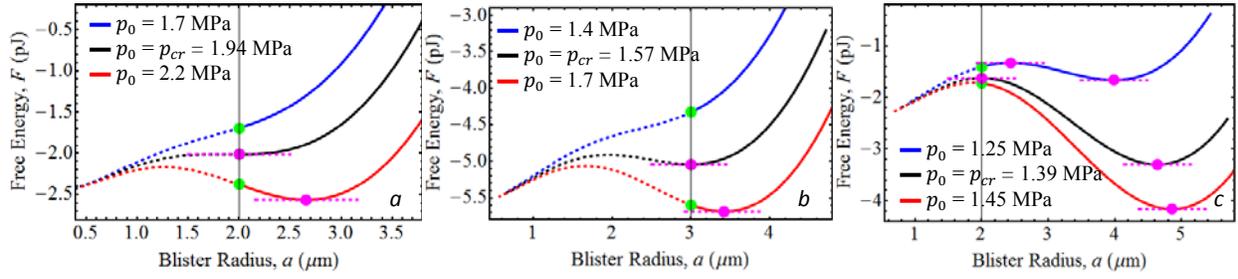

Figure 2  Variation of free energy with blister radius, at a fixed pressure $p_0$ with (a) $a_0 = 2$ $\mu m$ and $h = 0.25$ $\mu m$, (b) $a_0 = 3$ $\mu m$ and $h = 0.25$ $\mu m$ and (c) $a_0 = 2$ $\mu m$ and $h = 1.25$ $\mu m$

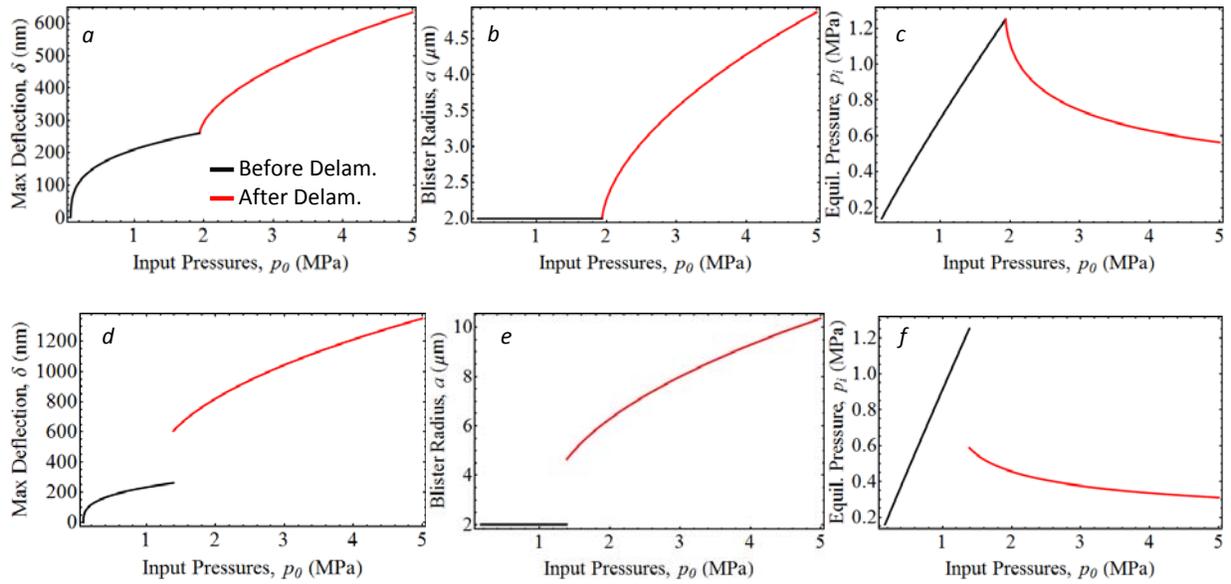

Figure 3     (a, d) Maximum deflection, $\delta$, (b, e) blister radius, $a$, and (c, f) final equilibrium microchamber pressure, $p_i$ plotted as functions of the input pressure, $p_0$ with $\Gamma = 0.2$ J/m$^2$. The cavity dimensions are (a-c) $a_0 = 2$ $\mu m$ and $h = 0.25$ $\mu m$ and (d-f) $a_0 = 2$ $\mu m$ and $h = 1.25$ $\mu m$

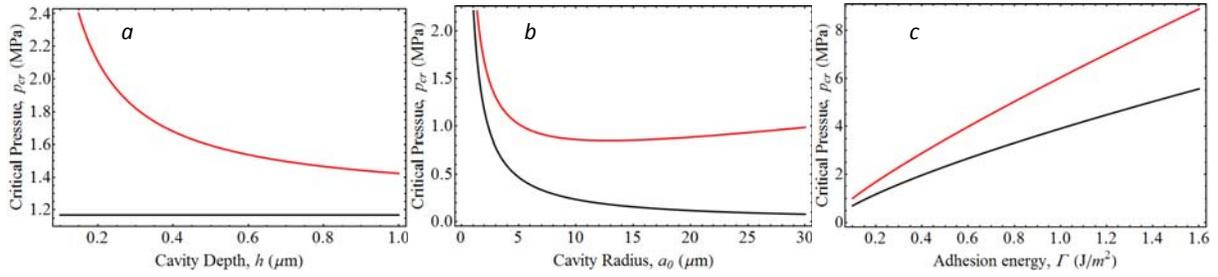

Figure 4   Critical pressure for the onset of delamination as a function of: (a) cavity depth, (b) cavity radius and (c) adhesion energy for the constant pressure (black curves) and constant N (red curves) blister tests. When not being varied, $h = 400$ $nm$, $a_0 = 2$ $\mu m$, and $\Gamma = 0.2$ J/m$^2$.

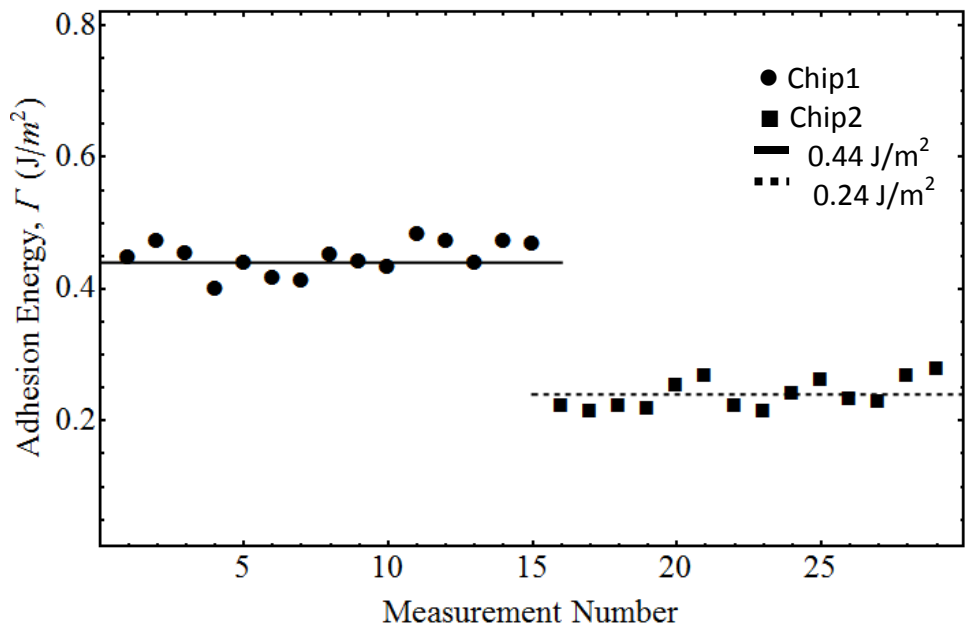

Figure 5   Adhesion energies for monolayer graphene membranes on two different $SiO_2$ substrates/chips. The average adhesion energy is 0.44 $J/m^2$ for Chip 1 and 0.24 $J/m^2$ for Chip 2.

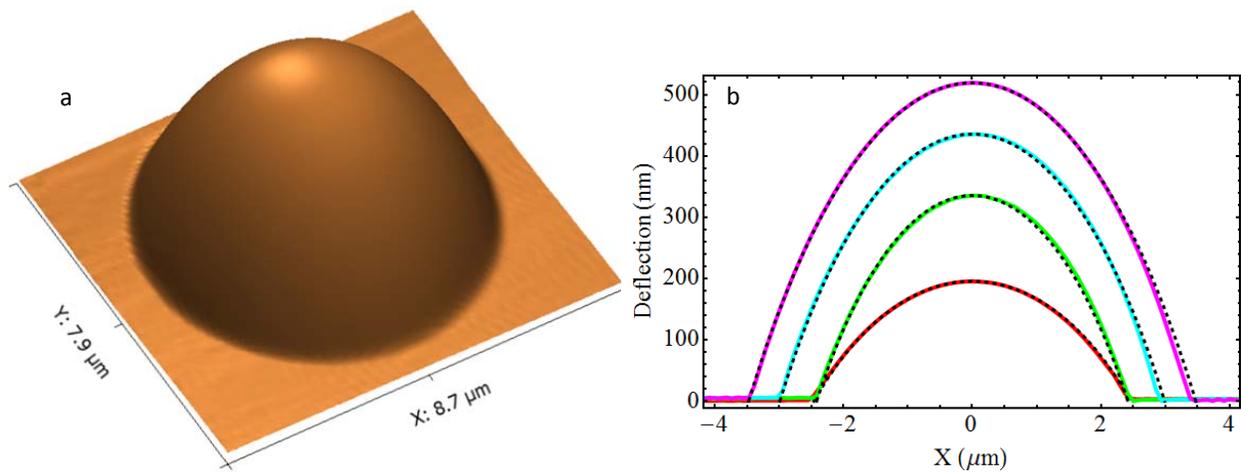

Figure 6  (a) Three dimensional rendering of AFM height scan of a graphene blister pressurized to 2.4 MPa (Chip 2). The maximum height is about 520 nm; (b) cross sections of the AFM height measurements (Chip 2) at different input pressures,     – 0.48 MPa (red), 1.32 MPa (green), 1.83 MPa (cyan) and 2.40 MPa (magenta). The black curves are the deflection profiles from Hencky's solution, with the maximum deflection fit to the measured value.

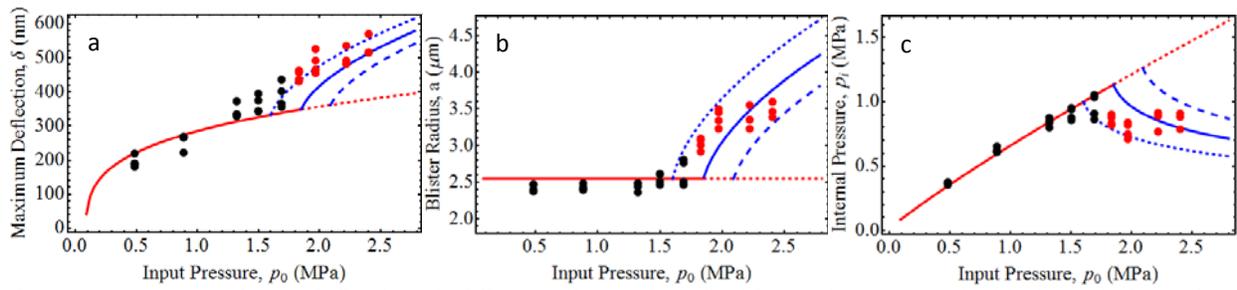

Figure 7 (a) Maximum deflection, (b) blister radius, and (c) final internal pressure. The black symbols are from measurements and the solid curves are from the analysis with: no delamination (red), and delamination for different values of adhesion energy: $\Gamma = 0.2$ J/m$^2$ (dashed blue), $\Gamma = 0.24$ J/m$^2$, and $\Gamma = 0.28$ J/m$^2$ (long dashed blue). The red symbols are those that were used to determine the adhesion energies in Fig. 5.

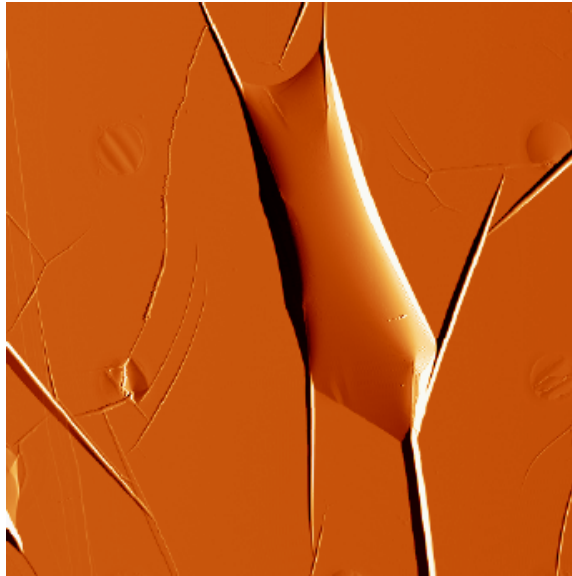

Figure 8  (a) AFM amplitude image (40×40 $\mu m$) of a graphene membrane that has undergone large-scale delamination at $p_0 = 2.8$ MPa with $a_0 \approx 2.2$ $\mu m$ and $h \approx 5$ $\mu m$. Assuming the adhesion energy is between 0.2-0.4 J/m$^2$ and the graphene has 8 layers, the critical pressure is between 1.2-1.9 MPa.